# Electrical Contacts to Three-Dimensional Arrays of Carbon Nanotubes

Aron W. Cummings, Julien Varennes, and François Léonard

*Abstract*—We use numerical simulations to investigate the properties of metal contacts to three-dimensional arrays of carbon nanotubes (CNTs). For undoped arrays top-contacted with high or low work function metals, electrostatic screening is very strong, resulting in a small Schottky barrier for current injection in the top layer and large Schottky barriers for current injection in the deeper layers. As a consequence, the majority of the current flows through the top layer of the array. Our simulations show that doping of the CNT array can alleviate this problem, even without direct contact to each tube in the array; however, we find that the charge transfer length is unusually long in arrays and increases with the number of CNT layers under the contact. We also show that a bottom gate can modulate the contact resistance, but only very weakly. These results are important for the design of electronic and optoelectronic devices based on CNT arrays, because they suggest that increasing the thickness of the array does little to improve the device performance unless the film is strongly doped at the contacts and the contact is long, or unless each tube in the array is directly contacted by the metal.

*Index Terms*—Carbon nanotubes, contact resistance, nanocontacts, nanotube devices

## I. Introduction

OWING to their unique electrical and optical properties, carbon nanotubes (CNTs) have emerged as a promising material for next-generation nanoelectronic and nanophotonic devices. Indeed, a variety of studies have demonstrated the potential of transistors [1],[2], diodes [3],[4], and photodetectors [5],[6] where an individual CNT is the active element. However, technology applications can require a higher density of CNTs. In nanoelectronic devices this would increase the total current, while for nanophotonic devices this would also increase the absorption cross section. Disordered CNT films offer a potential solution, but charge transport is often dominated by hopping between nanotubes, which severely limits the current [7]. An alternative is to use arrays of aligned CNTs. There have been several devices made from single layers of aligned CNTs [8]-[10], and it has been shown that the current scales with the number of tubes in the array. In an attempt to further improve performance, recent experimental work has examined CNT array devices that are much thicker than a single layer [11], i.e. three-dimensionally organized arrays.

While CNTs exhibit favorable properties, it is often the CNT-metal contact that governs the device performance. For example, the type of metal and the nanotube radius can determine whether the contact is ohmic or Schottky [12]-[14], and the contact geometry also plays a significant role [15]-[17]. In array devices, the presence of neighboring nanotubes becomes important, and electrostatic screening can modify the contact properties in single-layer CNT array transistors [18]. For thicker CNT arrays, the situation becomes more complicated. This is seen in Fig. 1, which depicts the system under consideration: a three-dimensional CNT array contacted with metal from above. In this geometry the metal only touches the top layer of CNTs, which raises some fundamental questions about the contact – is it ohmic or Schottky, and what is the nature of the charge injection between the metal and the array?

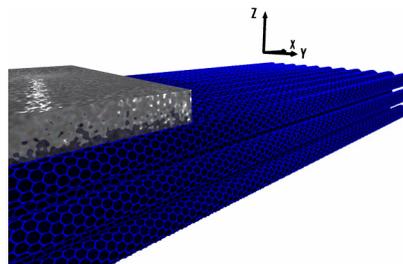

Fig. 1. Schematic of a top-contacted three-dimensional CNT array.

In this work, we use numerical simulations to study contacts to multilayer CNT arrays. We find that electrostatic screening within the array plays a fundamental role in determining the contact properties. Specifically, contacts to undoped arrays

Manuscript received June 3, 2013. This project is supported by the U.S. Department of Energy, Office of Science, through the National Institute for Nano-Engineering (NINE) at Sandia National Laboratories, and by the Laboratory Directed Research and Development program at Sandia National Laboratories, a multiprogram laboratory operated by Sandia Corporation, a Lockheed Martin Co., for the United States Department of Energy under Contract No. DEAC01-94-AL85000.

A. W. Cummings was with Sandia National Laboratories, MS9161, Livermore, CA 94551, USA. He is now with ICN2 - Institut Català de Nanociència i Nanotecnologia, Campus UAB, 08193 Bellaterra (Barcelona), Spain (e-mail: aron.cummings@icn.cat).

J. Varennes was with Sandia National Laboratories, MS9161, Livermore, CA 94551, USA. He is now with the Department of Physics, Purdue University, West Lafayette, IA 47907 USA (e-mail: julien.var@gmail.com).

F. Léonard is with Sandia National Laboratories, MS9161, Livermore, CA 94551 USA (e-mail: fleonar@sandia.gov).

lead to charge transfer that is limited primarily to the top layer of CNTs. This results in a small Schottky barrier for the top layer and large Schottky barriers for the deeper layers, implying that the majority of the current passes only through the top layer of the array. We show that to take advantage of all the CNTs in the array, strong doping is required to reduce the Schottky barrier for the deeper layers. However, even in the doped case, we find that the charge transfer length is unusually long and increases with the number of CNT layers under the contact. We also show that a bottom gate can be used to improve the contact resistance, but only very weakly due to the large electrostatic screening. These conclusions are important for the design of electronic devices based on CNT arrays, because they suggest that the array thickness has little effect on device performance unless the contact is very long and strongly doped, or unless each CNT in the array is directly contacted by the metal.

## II. METHODS

To determine the properties of the contacts, we use an atomistic, tight-binding numerical approach [17],[19]. The first step is a self-consistent calculation of the charge and potential within the array. The potential is obtained from the charge through a 3D solution of Poisson's equation, $\nabla \cdot (\varepsilon \nabla V) = -\rho$, where $\rho$ is the charge density, $V$ is the electrostatic potential, and $\varepsilon$ is the spatially-dependent dielectric constant. We treat the metals in the device by imposing Dirichlet boundary conditions at the edges of the contacts, and we assume Neumann boundary conditions at the edges of the simulation space in the x-z plane shown in figure 1. The type of contact metal is determined by the difference between its work function and that of the CNTs, $\Delta \varphi = \varphi_{CNT} - \varphi_{metal}$. The value of $\Delta \varphi$ then determines the potential in the contact, assuming the reference potential is at the CNT mid-gap. The array sits on top of 90 nm of SiO$_2$. Except for the gate-dependent results, we assume a floating gate geometry by applying Neumann boundary conditions (zero electric field) at the bottom of the SiO$_2$ layer. To include a bottom gate, we change this to a Dirichlet boundary condition with the applied gate voltage. To simulate an array, we impose periodic boundary conditions along the y-axis. Poisson's equation is discretized with a finite element scheme and is solved numerically with a conjugate gradient algorithm. This yields a 3D potential profile, $V(x, y, z)$.

To calculate the charge due to the potential, we describe the electronic structure of each CNT using an atomistic tight-binding representation. A common approximation is to assume a uniform electrostatic potential for all of the atoms around the CNT at a given axial position, allowing for the transformation to a single-band mode-space Hamiltonian [20]. However, we have found that in arrays this is no longer a good approximation due to the large variation of the potential across the CNT diameter. Thus, we used an atomistic approach such that the charge for each atom of the CNT is calculated independently. The charge on the $i$th carbon atom in the array is obtained from the potential according to $\rho_i = e/2\pi \cdot \int \text{Im} G^<_{ii}(E) dE + \rho_{doping}$, where $e$ is the electron charge and $\rho_{doping}$ is due to extrinsic doping of the CNT array. $G^<$ is the electron correlation function, determined by applying the atomistic tight-binding Hamiltonian to the non-equilibrium Green's function (NEGF) formalism [21], where the potential on the $i$th carbon atom, $V(x_i, y_i, z_i)$, goes into the $i$th diagonal element of the Hamiltonian. The atomistic charge density is then mapped back to a 3D distribution with a Gaussian smearing of the charge around each carbon atom.

Once the charge and potential have been determined self-consistently, we can calculate quantities relevant to the contact properties. Due to charge transfer between the metal contact and the CNT array, the conduction and valence bands of each nanotube will shift relative to the Fermi level, and the size of this shift is given by the self-consistent potential. For a large-work-function metal such as palladium (Pd), the Schottky barrier height for hole transport is given by $\phi_{SB} = eV_{avg} + E_g/2$, where $E_g$ is the band gap of the CNT and $V_{avg}$ is the average of the self-consistent potential around a ring of carbon atoms in the contact region. The justification for using $V_{avg}$ to calculate the Schottky barrier is as follows: In response to a perturbing potential, the energy bands of a CNT can be written as $E_n = E_n^{(0)} + E_n^{(1)}$, where $E_n^{(0)}$ is the energy of the $n$th subband without the perturbing potential, and $E_n^{(1)}$ is the first-order correction to this energy. If we assume a transverse electric field, then the potential on the $i$th atom in a carbon ring of the CNT can be written as $V_i = V_{avg} + \tilde{v}_i$, where $V_{avg}$ is the average potential around the carbon ring and $\tilde{v}_i$ is the deviation from this potential, such that $\sum_i \tilde{v}_i = 0$. Using first-order time-independent perturbation theory, it can be shown that $E_n^{(1)} = eV_{avg}$. In other words, $V_{avg}$ contributes a rigid shift of the energy bands of the CNT and $\tilde{v}_i$ has no effect. Using more detailed calculations of the band structure and band gap of the CNT, we have found that the transverse field does not become important until it reaches a value of 0.8 V/nm. Since the largest value we see in our simulations is 0.1 V/nm, we can safely ignore the effect of a transverse field and use the average potential to calculate the Schottky barrier height.

To calculate the contact resistance, we consider the conductance through the CNT array. For short channel lengths the channel resistance vanishes and the conductance is determined entirely by the contacts, $G = G_C$, where $G = 4e^2/h \cdot \int T(E)[-df(E)/dE]dE$ with $f(E)$ the Fermi function and the transmission $T(E)$ is calculated with the NEGF formalism. This allows us to write the contact resistance as $R_C = 1/G$. For the transport simulations, we use leads composed of the CNT array and the contact metal, and the length of the leads is set using the layer doubling approach of López Sancho *et al* [22]. An important addition to the numerical approach is electronic coupling between neighboring CNTs, which allows carriers from the bottom layers to reach the electrodes. This is included through an off-diagonal hopping term in the Hamiltonian that couples adjacent atoms on adjacent tubes. Prior work has estimated the coupling to be small, around a few meV [23],[24]. Varying the intertube coupling from 0 to 300 meV has little effect on our electrostatic results, but has a strong effect on the transport, as will be discussed below. We also include coupling to the contact metal by adding a term $-i\Delta/2$ to the diagonal elements of the Hamiltonian that correspond to the carbon atoms closest to the contact [17],[25],[26].

For our simulations, we consider square arrays of CNTs contacted by two Pd electrodes, which are known to form ohmic contacts to individual CNTs [1],[12]. The Pd work function is taken to be 1 eV larger than the CNT work function [12]. The array consists of semiconducting (17,0) CNTs, each with a diameter of 1.3 nm and a band gap of 0.54 eV within our tight-binding model. We report the contact resistance due to a single column of CNTs because the array is translationally invariant along the y-axis. Also, $R_C$ refers to the two-terminal contact resistance, i.e. the resistance measured in a transport experiment. This is related to the resistance of the left and right contacts by $R_C = R_C^L + R_C^R$.

### III. RESULTS AND DISCUSSION

We first discuss the case of infinitely long contacts. In Fig. 2(a) we show a schematic of a CNT array that is ten layers deep. In Fig. 2(b) we plot the conduction and valence band edges of the CNTs in each layer beneath the contact, for an intertube spacing of 0.35 nm. For the top layer (layer 1, in direct contact with the metal) the Fermi level sits close to the valence band edge. However, for deeper layers the Fermi level is deeper in the band gap, creating a Schottky barrier for holes, $\varphi_{SB}$. In Fig. 2(c) we plot the Schottky barrier height in each layer for intertube spacings from 0.35 to 10 nm. In this figure, two important features can be identified. The first is that only the top layer of the array has a small Schottky barrier $(\varphi_{SB} < kT)$. This small barrier originates from charge transfer between the metal and the top CNT layer [12]. The larger barriers in the deeper layers indicate that electrostatic screening is strong, such that the top layer effectively shields the rest of the array from the contact. One can also see that the Schottky barrier height decreases with increasing intertube spacing, but only weakly. Even at a spacing of 10 nm it is still only the top layer that has a small Schottky barrier, indicating that field penetration between the CNTs is not important at these densities.

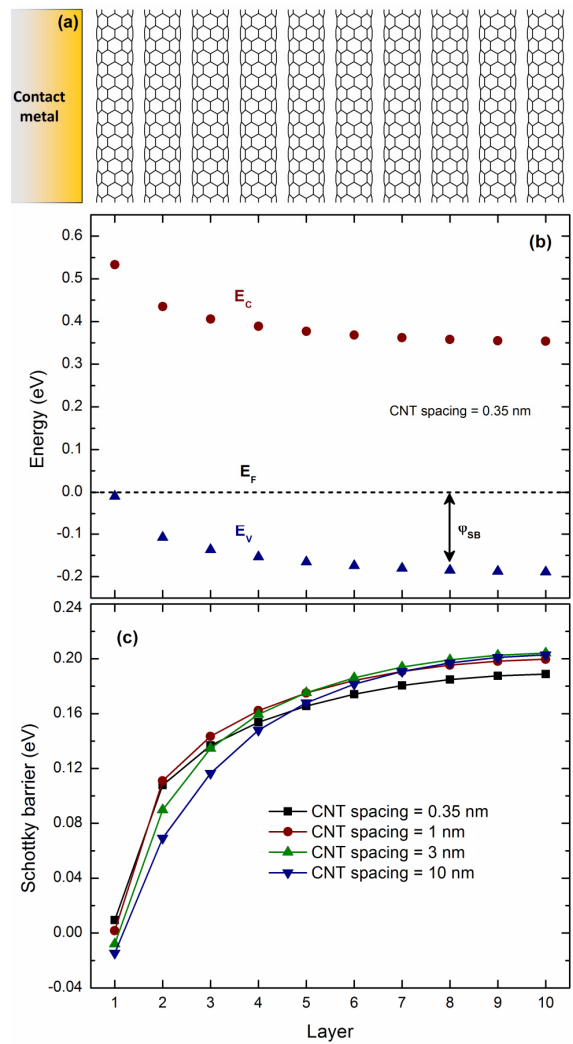

Fig. 2. Electrostatic characteristics of CNT array contacts. Part (a) shows a schematic of the CNT array, part (b) shows the position of the conduction and valence band edges of the CNTs in each layer, and part (c) shows the corresponding Schottky barrier height of the CNTs in each layer.

In order to examine the contact resistance, we turn to the results of the transport calculations. In Fig. 3(a) we plot the conductance (left axis) and resistance (right axis) of the nanotubes in each layer of the array. As seen in this figure, the conductance through the top layer is significantly larger than in the deeper layers. Similarly, the resistance of the CNTs in the top layers is significantly smaller than in the deeper layers. In Fig. 3(b) we plot the total conductance (left axis) and total contact resistance (right axis) as a function of array thickness. As the thickness increases from one to 10 layers, the total conductance only increases by 7%, and the contact resistance drops by the same amount. This result indicates that transport through the array is dominated by the top layer of CNTs, and that increasing the thickness beyond a single layer has little effect on the total contact resistance.

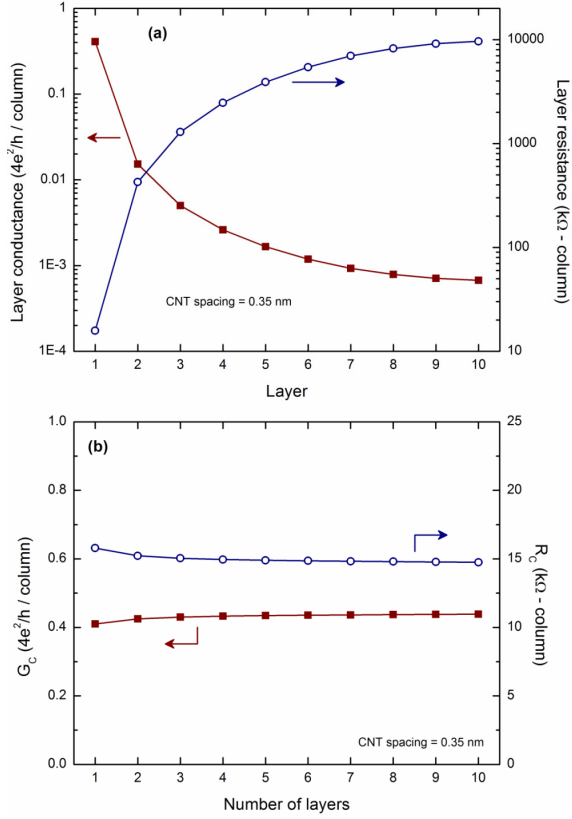

Fig. 3. Transport characteristics of the CNT array with infinitely long contacts. Part (a) shows the conductance and resistance of the nanotubes in each layer of the array. Part (b) shows the total conductance and the contact resistance of an array with thickness varying from one to 10 layers.

As suggested in Fig. 3, our transport calculations reveal an important result – the total conductance is equal to the sum of the individual conductances of each CNT in the array, $G = \sum_j G_j$. This can be understood as follows: although the contacts only couple to the top layer, the presence of intertube coupling means that if the contacts are long enough then charge carriers can hop from tube to tube, and all the CNTs in the array will contribute to transport, each with a conductance determined by its Schottky barrier. Assuming flat bands and ballistic transport, the conductance through an individual CNT is

$$G_j = 4e^2/h \cdot \left[1 + f\left(E_g^j - \phi_{SB}^j\right) - f\left(-\phi_{SB}^j\right)\right] \quad (1)$$

where $E_g^j$ and $\varphi_{SB}^j$ refer to the band gap and Schottky barrier height of the $j$th CNT in the array. The total contact resistance then becomes

$$R_C = \left(\sum_j G_j\right)^{-1} = \left(\sum_j 1/R_j\right)^{-1}. \quad (2)$$

These equations indicate that the array can be treated as a collection of isolated nanotubes, all conducting in parallel, and the transport through each nanotube is determined by its Schottky barrier height (in the ballistic regime). Furthermore, the total contact resistance is simply the parallel combination of all the CNTs in the array. Therefore, for undoped films transport is dominated by the top layer, and increasing the thickness beyond a single layer does little to improve the electrical properties of the array.

One question that arises is whether one can improve the contact properties by decreasing the Schottky barrier in the lower layers of the array. In traditional semiconductors, the usual way to decrease the contact resistance is with heavy doping near the contacts. To examine the effect of doping on contact resistance, we included uniform doping of the CNTs in our simulations. Fig. 4(a) shows the Schottky barrier height of each layer of the array for different doping densities, from 0 to $10^{-3}$ holes per carbon atom. One can see that doping has a strong effect on the Schottky barrier height of the deeper layers in the array. In particular, for a doping of $10^{-3}$ holes / C-atom the Schottky barrier height is nearly zero for all layers. Fig. 4(b) depicts the contact resistance as a function of doping density, and it is evident that doping can strongly reduce the overall contact resistance. Additionally, for strong doping the resistance of the deeper layers approaches that of the top layer, leading to an overall contact resistance that is reduced by the number of layers in the array.

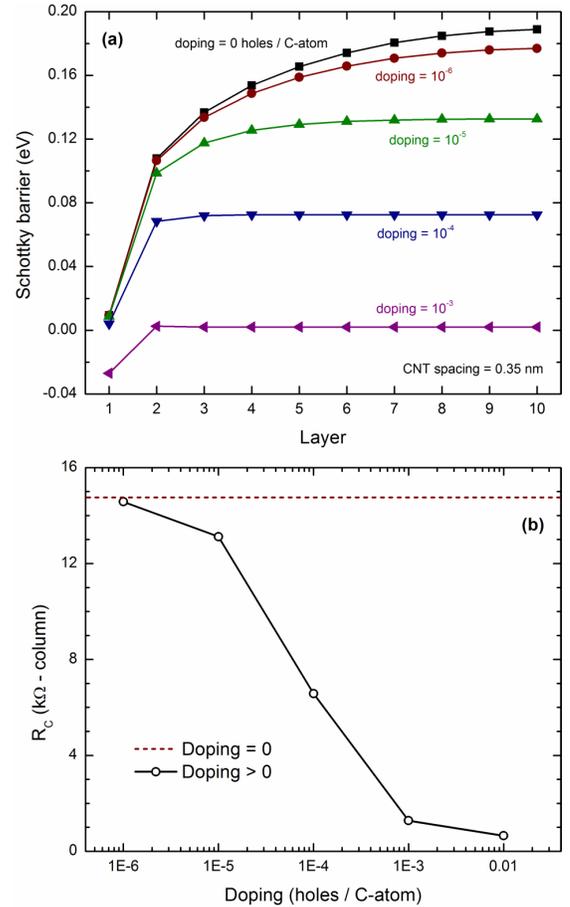

Fig. 4. Impact of doping on contacts to CNT arrays. Part (a) shows the Schottky barrier height of the CNTs in each layer at different doping densities. In part (b), the solid line shows the total contact resistance vs. doping density. The dashed line shows the contact resistance for zero doping.

Recently, several studies have shown that the contacts of CNT and graphene-based devices can be modulated by a bottom gate [17],[27],[28]. Therefore, we also consider the effect of a bottom gate on the contact properties, as shown in Fig. 5. In Fig. 5(a) we plot the Schottky barrier of each layer of CNTs for gate voltages from 0 to -5 V, with no doping included. In this plot, we see that a bottom gate can reduce the Schottky barrier height of the lower layers, but due to the strong electrostatic screening it has only a small effect on the middle layers of the array. In Fig. 5(b) we plot the contact resistance as a function of gate voltage, where the solid line is the total contact resistance and the dashed line is the resistance of the top layer only. This plot shows that because of the strong screening the bottom gate has only a marginal effect on the contact resistance of the array, and the total resistance is dominated by the top and bottom layers.

The results of Fig. 5 can be used to inform the design of transistors based on CNT arrays. Since the bottom gate has little to no effect on the middle and upper layers, the transistor would be less sensitive to the gate voltage, resulting in a device with a poor subthreshold swing. Putting the gate on top of the array would allow for direct modulation of the top layer of CNTs, giving much better gate control.

Our results also allow us to infer the behavior of more complicated contact geometries. For example, if the metal were to cover the top and sides of the array, the same screening would occur and only the CNTs on the outer layers of the array would have good contact. Slightly better contact might arise at the corners of the array, but for the macro- and mesoscopic films that we consider this would only be a small fraction of the conductivity. In general, to take advantage of the entire array, direct contact to each nanotube will be necessary.

It should be noted that these simulations have considered ideal arrays consisting entirely of semiconducting CNTs. However, real CNT arrays are likely to be populated with a mixture of semiconducting and metallic nanotubes. In this case, the presence of metallic CNTs would enhance the electrostatic screening within the array [18]. This would frustrate gate modulation of the semiconducting CNTs in the array and would also enhance the already strong electrostatic screening that we see under the contacts. Thus, the presence of metallic CNTs would further strengthen our conclusion that conduction is limited primarily to the top layer of the array.

While it appears that using a bottom gate is a poor way to modulate the contact properties, the doping results in Fig. 4 are encouraging because they indicate that it is possible for all the CNTs in the array to contribute to electrical transport. However, our simulations included a significant approximation – the use of semi-infinite leads for the contacts. In real devices the contacts are finite in length, and thus it is important to study the relationship between contact length and contact resistance in CNT arrays. This was studied experimentally for individual CNTs [2],[29] and graphene [30], where it was shown that the contact resistance decreases with increasing contact length and saturates for lengths on the order of several hundred nanometers. This saturation length is determined by the charge transfer length, a measure of how far a carrier will move under the contact before injection into the metal [31]. For individual CNTs, theory has shown that the charge transfer length is inversely proportional to the strength of the CNT-metal interaction [25].

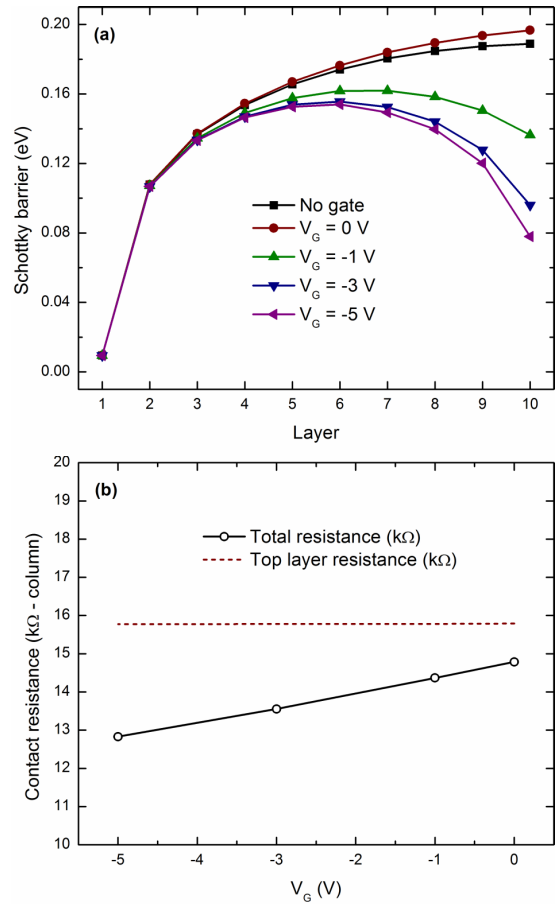

Fig. 5. Gate-dependent properties of contacts to CNT arrays. Part (a) shows the Schottky barrier height of the CNTs in each layer at different gate voltages. In part (b) the solid line shows the contact resistance as a function of gate voltage. The dashed line shows the resistance for the top layer only.

In Fig. 6 we plot the contact resistance of the CNT array as a function of contact length for a doping of $10^{-3}$ holes / C-atom. In Fig. 6(a) we assume an intertube coupling of 60 meV, equal to the CNT-metal coupling. In Fig. 6(b) we use a coupling of 300 meV, typical of graphene bilayers [32]. The solid lines show the contact resistance for films of varying thickness, and the dashed line shows the contact resistance of a four-layer film assuming an ideal situation with immediate charge hopping between CNTs. For all thicknesses, the contact resistance decays toward a constant value with increasing contact length. However, the length scale and nature of that decay depend strongly on the number of layers and on the coupling between CNTs. For short contacts (< 100 nm) the resistance is equal to the single-layer case, independent of array thickness, so little is gained by having multiple layers even with the high doping. For contact lengths around 100 nm, the contact resistance deviates from the single-layer array and eventually saturates at a value dependent on the number of layers according to (2), i.e. the resistance is reduced by the number of layers in the array. One can also see that the charge transfer length increases significantly with the number of layers. In Fig. 6(a) the charge transfer length increases from under 100 nm for a single-layer array up to 20 $\mu$m for a four-layer array. Even for relatively strong interlayer coupling, as in Fig. 6(b), the charge transfer length is almost 500 nm for a four-layer array.

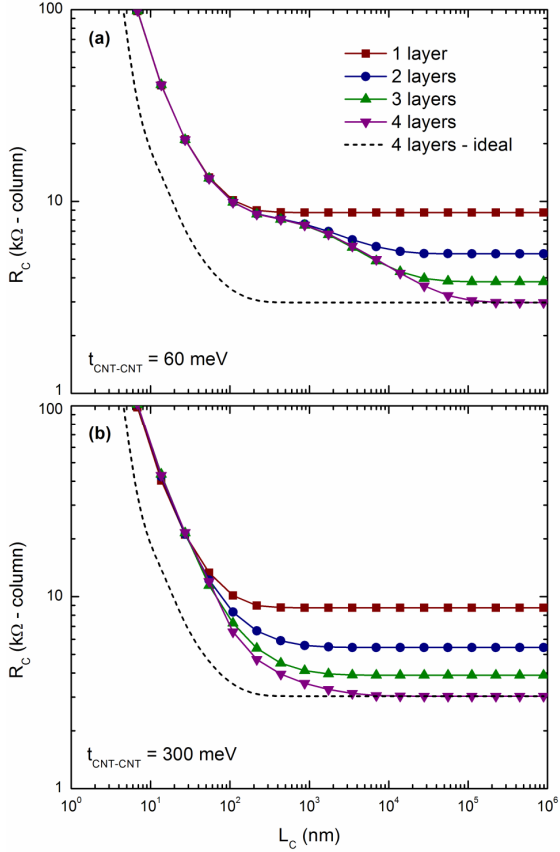

Fig. 6. Contact resistance vs. contact length for doped CNT arrays. The panels correspond to (a) intermediate and (b) strong intertube coupling. In both panels the dashed line is for an ideal four-layer array.

These results can be understood using an analytical model of transport through multilayer arrays. For simplicity we consider a two-layer array. In this model the resistance of the top layer is $R_0$, and the resistance of the second layer is $\beta R_0$, where $\beta > 0$. We also assume a conductance per unit length, $g$, for current injection between the top layer and the contact, and assume the charge injection between the layers is $\alpha g$, where $\alpha > 0$. In this model, a large value of $\alpha$ corresponds to large intertube coupling. Following the approach of Solomon [31], the contact resistance of the two-layer system is [33]

$$R_C(L_C) = \frac{R_0}{P_+ \tanh(L_C/L_+) + P_- \tanh(L_C/L_-)} \quad (3)$$

where $P_\pm$ and $L_\pm$ depend on $\alpha$, $\beta$, $g$, and $R_0$. Fig. 7 shows the contact resistance for this model using $\beta = 1$, $L_0 = 2/gR_0 = 10$ nm, and $R_0 = h/4e^2$. As the intertube coupling is increased, the model reproduces the transition from two plateaus to a single plateau as observed in the numerical simulations in Fig. 6. The presence of two plateaus at small coupling can be understood by considering (3) for $\beta = 1$ in the limit $\alpha \ll 1$. In that case, (3) becomes

$$R_C(L_C) = \frac{R_0}{\tanh(L_C/L_0) + \tanh(\alpha L_C/L_0)} \quad (4)$$

where $L_0 = 2/gR_0$ is the charge transfer length for the single-layer system. Thus, for weak coupling the contact resistance follows that of the top layer until a contact length of $L_0/\alpha$ is reached. At this point the contact resistance decreases due to the contribution of the second layer.

For strong intertube coupling, when $\alpha \gg 1$, (3) becomes

$$R_C(L_C) = \frac{R_0/2}{\tanh(L_C/2L_0)} . \quad (5)$$

Equation (5) has the form of a single-layer contact resistance with a single plateau, but with an important difference: the charge transfer length, $2L_0$, is double that of the single-layer case despite the strong interlayer coupling. Within our model, the rate of charge transfer between layers is proportional to the voltage drop between them. For a given bias, the voltage drop across neighboring layers decreases as the number of layers increases and thus the rate of charge transfer also decreases. This result is important because it sets a lower bound on the charge transfer length of multilayer arrays. Specifically, for an $N$-layer array in the strong coupling limit the charge transfer length is $NL_0$, $N$ times greater than the single-layer case.

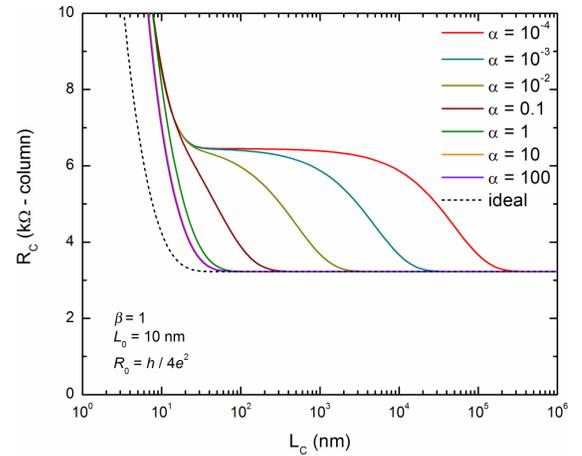

Fig. 7. Contact resistance vs. contact length of a two-layer array, calculated with the analytical model described in the main text. The curve for $\alpha = 10$ is not visible because it overlaps with the curve for $\alpha = 100$.

To appreciate the relationship between $\alpha$ and the interlayer coupling, we can fit the analytical model to our numerical results. Specifically, we fitted (3) to the two-layer numerical results in Fig. 6 [33]. In this case an interlayer coupling of 60 meV corresponds to $\alpha = 0.009$, while an interlayer coupling of 300 meV corresponds to $\alpha = 0.28$. Thus, even for relatively strong interlayer coupling, $\alpha$ remains rather small and we are far from the lower bound described by (5). These results suggest that top-contacted multilayer arrays suffer from a fundamental issue – due to the cascading effect of charge transfer between CNT layers, the contacts must be very long to realize the benefit of having multiple layers in the array.

IV. CONCLUSION

In summary, we used numerical simulations to investigate the properties of top contacts to three-dimensional CNT arrays. We find that screening confines charge transport to the top layer unless the CNTs are strongly doped. This strong screening also results in inefficient gating of the array. In addition, we find that the charge transfer length can be very long even for strong intertube coupling. These results suggest that direct contact to each nanotube in the array may be necessary to achieve the ultimate benefits of such systems, especially if short contacts are needed. These results can also motivate the design and fabrication of single-layer devices. Since transport is confined to the top layer of the array, there may be no need to precisely control the number of vertical layers, leading to more reliable and uniform operation from device to device. While the results focused on nanotube arrays, we expect that similar issues will arise in other three-dimensionally layered systems, including nanowires and graphene.

**Supplemental Material to "Electrical Contacts to Three-Dimensional Arrays of Carbon Nanotubes",**

**by A.W. Cummings, J. Varennes, and F. Léonard**

A. Derivation of the length-dependent contact resistance of a two-layer CNT array

To derive an analytical model of the two-layer contact resistance we consider the situation depicted in figure S1, which closely follows the approach of Solomon [S1]. We assume that the resistance of the top layer is $R_0$, and the resistance of the second layer is $\beta R_0$, where $\beta > 0$ is a real number. We also assume a conductance per unit length, $g$, for current injection between the top CNT layer and the metal contact, and assume the charge injection between the nanotube layers is given by $\alpha g$, where $\alpha > 0$ is real. The voltage on the contact is given by $V_C$, and the length of the contact is $L_C$. The position-dependent current in the contact is made up of forward-going and reverse-going components in each layer, labeled $I_1^{f,r}(x)$ and $I_2^{f,r}(x)$, respectively. Here, a forward-going (reverse-going) current represents a current moving in the +x (-x) direction.

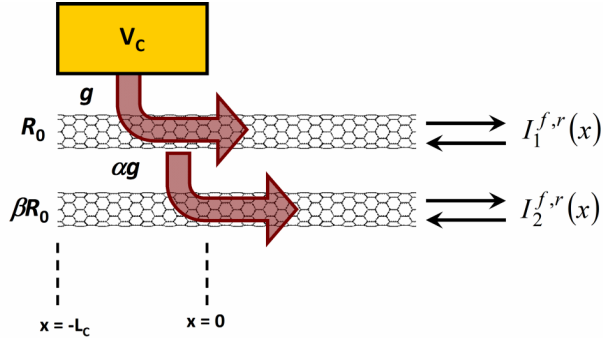

Fig. S1. Model of a two-layer CNT array used for the contact resistance calculation.

The current and the voltage of the forward- and reverse-going streams in each layer can be related through Ohm's law,

$$I_{1,2}^{f,r} = V_{1,2}^{f,r}/R_{1,2}, \tag{S1}$$

where $R_1 = R_0$ and $R_2 = \beta R_0$ from figure S1. The total current in each layer is given by the difference of the forward- and reverse-going streams,

$$I_{1,2} = I_{1,2}^{f} - I_{1,2}^{r}. \tag{S2}$$

The net voltage in each layer is given by the average voltage due to the forward- and reverse-going streams of current,

$$V_{1,2} = \left(V_{1,2}^{f} + V_{1,2}^{r}\right)/2. \tag{S3}$$

In the top layer, current can be transferred to/from the contact metal and to/from the bottom layer, while in the bottom layer current is transferred to/from the top layer. The rate of current exchange depends on the voltage difference between the layers and also on the interlayer conductance,

$$\frac{dI_1^{f,r}}{dx} = \pm \frac{1}{2} g \left( V_C - V_1^{f,r} \right) \pm \frac{1}{2} \alpha g \left( V_2^{f,r} - V_1^{f,r} \right) \tag{S4}$$

and

$$\frac{dI_2^{f,r}}{dx} = \mp \frac{1}{2} \alpha g \left( V_2^{f,r} - V_1^{f,r} \right). \tag{S5}$$

To simplify things, we can subtract the forward- and reverse-going components of equations (S4) and (S5), and use equations (S2) and (S3) to give

$$\frac{dI_1}{dx} = g \left( V_C - V_1 \right) + \alpha g \left( V_2 - V_1 \right) \tag{S6}$$

and

$$\frac{dI_2}{dx} = -\alpha g \left( V_2 - V_1 \right). \tag{S7}$$

Next, we can substitute equation (S1) into equations (S4) and (S5), add the forward- and reverse-going components, and use equations (S2) and (S3) to get

$$\frac{dV_1}{dx} = -\frac{\alpha+1}{2} \frac{R_0}{L_0} I_1 + \frac{\alpha\beta}{2} \frac{R_0}{L_0} I_2 \tag{S8}$$

and

$$\frac{dV_2}{dx} = \frac{\alpha\beta}{2} \frac{R_0}{L_0} I_1 - \frac{\alpha\beta^2}{2} \frac{R_0}{L_0} I_2, \tag{S9}$$

where $L_0 = 2/gR_0$ is the charge transfer length between the top layer and the contact metal. Finally, we take the derivative of equations (S6) and (S7) and substitute in equations (S8) and (S9) to arrive at

$$\begin{bmatrix} \dfrac{d^2 I_1}{dx^2} \\ \dfrac{d^2 I_2}{dx^2} \end{bmatrix} = \frac{1}{L_0^2} \begin{bmatrix} \alpha^2 \beta + (\alpha+1)^2 & -\alpha\beta(1+\alpha+\alpha\beta) \\ -\alpha(1+\alpha+\alpha\beta) & \alpha^2 \beta(\beta+1) \end{bmatrix} \begin{bmatrix} I_1(x) \\ I_2(x) \end{bmatrix}$$

$$= \frac{1}{L_0^2} \begin{bmatrix} A_{11} & -A_{12} \\ -A_{21} & A_{22} \end{bmatrix} \begin{bmatrix} I_1(x) \\ I_2(x) \end{bmatrix}. \tag{S10}$$

The general solution of equation (S10) is given by

$$\begin{bmatrix} I_1(x) \\ I_2(x) \end{bmatrix} = c_1 \begin{bmatrix} 1 \\ d_+ \end{bmatrix} e^{x/L_+} + c_2 \begin{bmatrix} 1 \\ d_+ \end{bmatrix} e^{-x/L_+} + c_3 \begin{bmatrix} 1 \\ d_- \end{bmatrix} e^{x/L_-} + c_4 \begin{bmatrix} 1 \\ d_- \end{bmatrix} e^{-x/L_-},$$

where

$$L_\pm = L_0 \sqrt{\frac{2}{(A_{11} + A_{22}) \pm \sqrt{(A_{11} + A_{22})^2 - 4(A_{11}A_{22} - A_{12}A_{21})}}} \quad \text{(S11)}$$

and

$$d_\pm = \frac{A_{11} - L_0^2/L_\pm^2}{A_{12}}.$$

Using the boundary condition that $I_{1,2}(-L_C) = 0$, equation (S11) can be rewritten as

$$\begin{bmatrix} I_1(x) \\ I_2(x) \end{bmatrix} = c_1 \begin{bmatrix} 1 \\ d_+ \end{bmatrix} \left( e^{x/L_+} - e^{-(x+2L_C)/L_+} \right) + c_3 \begin{bmatrix} 1 \\ d_- \end{bmatrix} \left( e^{x/L_-} - e^{-(x+2L_C)/L_-} \right). \quad \text{(S12)}$$

To solve for $c_1$ and $c_3$, we take the derivative of equation (S12), equate it to equations (S6) and (S7), and use the boundary condition $V_{1,2}(0) = 0$. This yields

$$c_{1,3} = \pm \frac{2V_C}{R_0} \cdot \frac{L_\pm}{L_0} \cdot \frac{1/d_\pm}{1/d_+ - 1/d_-} \cdot \frac{1}{1 + e^{-2L_C/L_\pm}}. \quad \text{(S13)}$$

Finally, the length-dependent contact resistance of the two-terminal device can be written as $R_C = 2V_C/I_{tot}(0)$, where $I_{tot}(x) = I_1(x) + I_2(x)$. This yields

$$R_C(L_C) = \frac{R_0}{P_+ \tanh(L_C/L_+) + P_- \tanh(L_C/L_-)},$$

where $\quad \text{(S14)}$

$$P_\pm = \frac{L_\pm}{L_0} \cdot \frac{1 + 1/d_\pm}{1/d_+ - 1/d_-}.$$

B. Fit of analytical model to numerical results

In order to fit equation (S14) to the numerical results shown in figure 6 of the main text, we must find appropriate values for $L_0$, $R_0$, $\beta$, and $\alpha$. Since we know the Schottky barrier heights of the top two layers of the array (see figure 4 of the main text), we can calculate $R_0$ and $\beta$ by inverting equation (1) of the main text. Next, we find $L_0$ by fitting the analytical form of the contact resistance of a single-layer array to the numerical results, where the single-layer contact resistance is given by

$$R_C(L_C) = \frac{R_0}{\tanh(L_C/L_0)}. \tag{S15}$$

Once $L_0$, $R_0$, and $\beta$ are known, we can adjust $\alpha$ to provide the best fit of equation (S14) to the two-layer numerical results shown in figure 6 of the main text. The results of this fit are shown in figure S2 below.

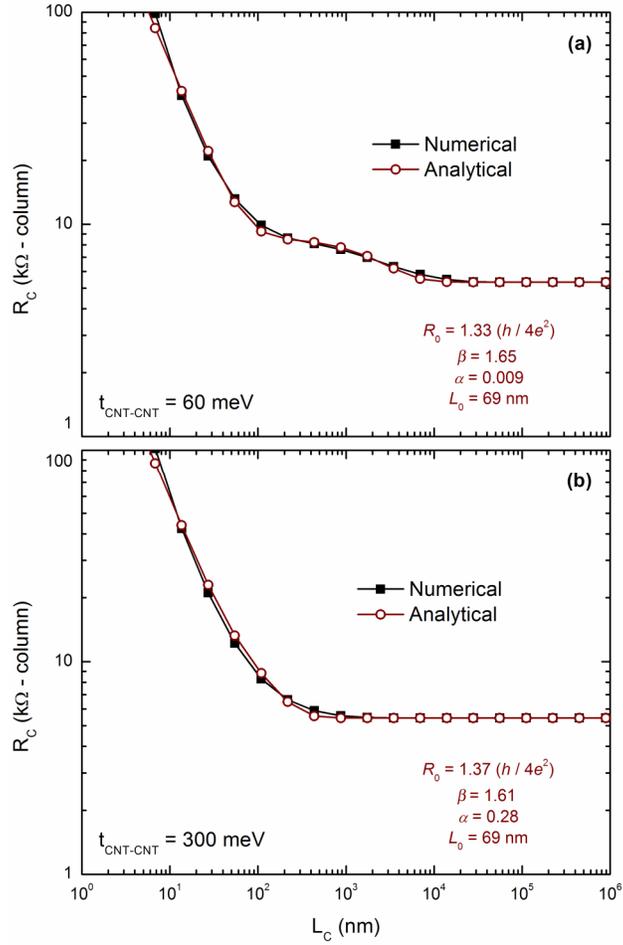

Fig. S2. Fit of the analytical contact resistance to the numerical results for a two-layer CNT array, with (a) intermediate and (b) strong interlayer coupling.